\documentclass[10pt,a4paper]{article}
\usepackage[export]{adjustbox}
\usepackage{amsmath,amssymb,mathrsfs,mathtools}
\usepackage[top=4cm,bottom=4cm,left=2.5cm,right=3.5cm]{geometry}
\usepackage{fancyhdr}
\usepackage{subcaption}
\usepackage{tikz}%
\newcommand{\RNum}[1]{\uppercase\expandafter{\romannumeral #1\relax}}
\usetikzlibrary{arrows.meta}
\usetikzlibrary{patterns}
\usetikzlibrary{decorations}
\usetikzlibrary{decorations.markings}
\usetikzlibrary{decorations.pathreplacing}
\usetikzlibrary{shapes,snakes}
\definecolor{blue-green}{rgb}{0.0,0.87,0.87}
\definecolor{caribbeangreen}{rgb}{0.0,0.8,0.6}
\definecolor{cornflowerblue}{rgb}{0.39,0.58,0.93}
\definecolor{darkspringgreen}{rgb}{0.09,0.45,0.27}
\definecolor{emerlad}{rgb}{0.31,0.78,0.47}
\definecolor{frenchblue}{rgb}{0.0,0.45,0.73}
\definecolor{frenchrose}{rgb}{0.96,0.29,0.54}
\definecolor{darkblue}{rgb}{0.0,0.0,0.55}
\definecolor{airforceblue}{rgb}{0.36, 0.54, 0.66}
\definecolor{amethyst}{rgb}{0.6, 0.4, 0.8}
\definecolor{awesome}{rgb}{1.0, 0.13, 0.32}
\definecolor{azure(colorwheel)}{rgb}{0.0, 0.5, 1.0}
\definecolor{ballblue}{rgb}{0.13, 0.67, 0.8}
\definecolor{blue}{rgb}{0.0, 0.0, 1.0}
\definecolor{blue(munsell)}{rgb}{0.0, 0.5, 0.69}
\definecolor{blue(ncs)}{rgb}{0.0, 0.53, 0.74}
\definecolor{blue(ryb)}{rgb}{0.01, 0.28, 1.0}
\definecolor{bostonuniversityred}{rgb}{0.8, 0.0, 0.0}
\definecolor{cadmiumred}{rgb}{0.89, 0.0, 0.13}
\definecolor{candyapplered}{rgb}{1.0, 0.03, 0.0}
\definecolor{crimson}{rgb}{0.86, 0.08, 0.24}%
\definecolor{darkpastelgreen}{rgb}{0.01, 0.75, 0.24}
\definecolor{darkslateblue}{rgb}{0.28, 0.24, 0.55}
\definecolor{dartmouthgreen}{rgb}{0.05, 0.5, 0.06}
\definecolor{deepjunglegreen}{rgb}{0.0, 0.29, 0.29}
\definecolor{denim}{rgb}{0.08, 0.38, 0.74}
\definecolor{dollarbill}{rgb}{0.52, 0.73, 0.4}
\definecolor{eggplant}{rgb}{0.38, 0.25, 0.32}
\definecolor{electriccyan}{rgb}{0.0, 1.0, 1.0}
\definecolor{electricgreen}{rgb}{0.0, 1.0, 0.0}
\definecolor{electricpurple}{rgb}{0.75, 0.0, 1.0}
\definecolor{gamboge}{rgb}{0.89, 0.61, 0.06}
\definecolor{gray-asparagus}{rgb}{0.27, 0.35, 0.27}
\definecolor{green(colorwheel)(x11green)}{rgb}{0.0, 1.0, 0.0}
\definecolor{green(html/cssgreen)}{rgb}{0.0, 0.5, 0.0}
\definecolor{green(pigment)}{rgb}{0.0, 0.65, 0.31}
\definecolor{green-yellow}{rgb}{0.68, 1.0, 0.18}
\definecolor{hotpink}{rgb}{1.0, 0.41, 0.71}
\definecolor{hotmagenta}{rgb}{1.0, 0.11, 0.81}
\definecolor{internationalorange}{rgb}{1.0, 0.31, 0.0}
\definecolor{jade}{rgb}{0.0, 0.66, 0.42}
\definecolor{islamicgreen}{rgb}{0.0, 0.56, 0.0}
\definecolor{lasallegreen}{rgb}{0.03, 0.47, 0.19}
\definecolor{lapislazuli}{rgb}{0.15, 0.38, 0.61}
\definecolor{lava}{rgb}{0.81, 0.06, 0.13}
\definecolor{lavendermagenta}{rgb}{0.93, 0.51, 0.93}
\definecolor{lawngreen}{rgb}{0.49, 0.99, 0.0}
\definecolor{lightcarminepink}{rgb}{0.9, 0.4, 0.38}
\definecolor{lightfuchsiapink}{rgb}{0.98, 0.52, 0.9}
\definecolor{lightseagreen}{rgb}{0.13, 0.7, 0.67}
\definecolor{lightsalmon}{rgb}{1.0, 0.63, 0.48}
\definecolor{limegreen}{rgb}{0.2, 0.8, 0.2}
\definecolor{lust}{rgb}{0.9, 0.13, 0.13}
\definecolor{malachite}{rgb}{0.04, 0.85, 0.32}
\definecolor{mediumslateblue}{rgb}{0.48, 0.41, 0.93}
\definecolor{mediumspringgreen}{rgb}{0.0, 0.98, 0.6}
\definecolor{mediumtealblue}{rgb}{0.0, 0.33, 0.71}
\definecolor{midnightblue}{rgb}{0.1, 0.1, 0.44}
\definecolor{midnightgreen(eaglegreen)}{rgb}{0.0, 0.29, 0.33}
\definecolor{mulberry}{rgb}{0.77, 0.29, 0.55}
\definecolor{oceanboatblue}{rgb}{0.0, 0.47, 0.75}
\definecolor{orange(ryb)}{rgb}{0.98, 0.6, 0.01}
\definecolor{pansypurple}{rgb}{0.47, 0.09, 0.29}
\definecolor{peridot}{rgb}{0.9, 0.89, 0.0}
\definecolor{persianindigo}{rgb}{0.2, 0.07, 0.48}%
\definecolor{persianblue}{rgb}{0.11, 0.22, 0.73}
\definecolor{pinegreen}{rgb}{0.0, 0.47, 0.44}
\definecolor{red(munsell)}{rgb}{0.95, 0.0, 0.24}
\definecolor{rosepink}{rgb}{1.0, 0.4, 0.8}
\definecolor{rosevale}{rgb}{0.67, 0.31, 0.32}
\definecolor{rosewood}{rgb}{0.4, 0.0, 0.04}
\definecolor{st.patrickblue}{rgb}{0.14, 0.16, 0.48}%
\definecolor{sangria}{rgb}{0.57, 0.0, 0.04}
\definecolor{scarlet}{rgb}{1.0, 0.13, 0.0}
\definecolor{seagreen}{rgb}{0.18, 0.55, 0.34}
\definecolor{shamrockgreen}{rgb}{0.0, 0.62, 0.38}
\definecolor{skobeloff}{rgb}{0.0, 0.48, 0.45}
\definecolor{skymagenta}{rgb}{0.81, 0.44, 0.69}
\definecolor{smokeytopaz}{rgb}{0.58, 0.25, 0.03}
\definecolor{teal}{rgb}{0.0, 0.5, 0.5}
\definecolor{tealgreen}{rgb}{0.0, 0.51, 0.5}
\definecolor{tiffanyblue}{rgb}{0.04, 0.73, 0.71}\definecolor{toolbox}{rgb}{0.45, 0.42, 0.75}
\definecolor{trueblue}{rgb}{0.0, 0.45, 0.81}
\definecolor{tropicalrainforest}{rgb}{0.0, 0.46, 0.37}
\definecolor{tuftsblue}{rgb}{0.28, 0.57, 0.81}
\definecolor{ultramarine}{rgb}{0.07, 0.04, 0.56}%
\definecolor{venetianred}{rgb}{0.78, 0.03, 0.08}
\definecolor{utahcrimson}{rgb}{0.83, 0.0, 0.25}%
\definecolor{vividcerise}{rgb}{0.85, 0.11, 0.51}
\definecolor{warmblack}{rgb}{0.0, 0.26, 0.26}
\definecolor{yaleblue}{rgb}{0.06, 0.3, 0.57}%
\definecolor{zaffre}{rgb}{0.0, 0.08, 0.66}%
\usepackage{xcolor}
\usepackage{hyperref}
\hypersetup{colorlinks, citecolor={green(html/cssgreen)},linkcolor={awesome}, urlcolor={zaffre}}

\begin{document}
\begin{titlepage}

\vskip 1.5in
\begin{center}
{\bf\Large{Mutation and Random Matrix Theory}}
\vskip
0.5cm {Mehdi Ameri\footnote{\color{blue}sth@um.ac.ir}\vspace{5pt}}\\{\small{ \textit{Department of Physics, Faculty of Science, Ferdowsi University of Mashhad\\ P.O.Box 1436, Mashhad, Iran\vspace{5pt}}}}
\end{center}
\vskip 0.5in
\baselineskip 16pt
\begin{abstract} We will study the relationship between two well-known theories, genetic evolution and random matrix theory in the context of many-body systems. It is suggested that genetic evolution can be described by a random matrix theory with statistical distribution in which mutation acts as a Gross-Witten-Wadia phase transition.
\end{abstract}

\end{titlepage}
\def\SO{{\mathrm{SO}}}
\def\G{{\text{\sf G}}}
\def\la{\langle}
\def\ra{\rangle}
\def\g{{\cmmib g}}
\def\U{{\mathrm U}}
\def\M{{\mathcal M}}
\def\d{{\mathrm d}}
\def\CS{{\mathrm{CS}}}
\def\Z{{\Bbb Z}}
\def\R{{\Bbb R}}
\def\J{{\mathcal J}}
\def\Bbb{\mathbb}
\def\Tr{{\rm Tr}}
\def\16{{\bf 16}}
\def\1{{(1)}}
\def\2{{(2)}}
\def\3{{\bf 3}}
\def\4{{\bf 4}}
\def\sg{{\mathrm g}}
\def\i{{\mathrm i}}
\def\h{\widehat}
\def\u{u}
\def\D{D}
\def\sp{{\sigma}}
\def\E{{\mathcal E}}
\def\O{{\mathcal O}}
\def\PF{{\mathit{P}\negthinspace\mathit{F}}}
\def\tr{{\mathrm{tr}}}
\def\be{\begin{equation}}
\def\ee{\end{equation}}
 \def\Sp{{\mathrm{Sp}}}
 \def\Spin{{\mathrm{Spin}}}
 \def\SL{{\mathrm{SL}}}
 \def\SU{{\mathrm{SU}}}
 \def\SO{{\mathrm{SO}}}
 \def\ll{\langle\langle}
\def\rr{\rangle\rangle}
\def\la{\langle}
\def\ra{\rangle}
\def\T{{\mathcal T}}
\def\V{{\mathcal V}}
\def\bar{\overline}
\def\v{v}

\def\tilde{\widetilde}
\def\t{\widetilde}
\def\R{{\Bbb{R}}}\def\Z{{\Bbb{Z}}}
\def\N{{\mathcal N}}
\def\B{{\mathcal B}}
\def\H{{\mathcal H}}
\def\hat{\widehat}
\def\Pf{{\mathrm{Pf}}}
\def\PSL{{\mathrm{PSL}}}
\def\Im{{\mathrm{Im}}}
\font\teneurm=eurm10 \font\seveneurm=eurm7 \font\fiveeurm=eurm5
\newfam\eurmfam
\textfont\eurmfam=\teneurm \scriptfont\eurmfam=\seveneurm
\scriptscriptfont\eurmfam=\fiveeurm
\def\eurm#1{{\fam\eurmfam\relax#1}}
\font\teneusm=eusm10 \font\seveneusm=eusm7 \font\fiveeusm=eusm5
\newfam\eusmfam
\textfont\eusmfam=\teneusm \scriptfont\eusmfam=\seveneusm
\scriptscriptfont\eusmfam=\fiveeusm
\def\eusm#1{{\fam\eusmfam\relax#1}}
\font\tencmmib=cmmib10 \skewchar\tencmmib='177
\font\sevencmmib=cmmib7 \skewchar\sevencmmib='177
\font\fivecmmib=cmmib5 \skewchar\fivecmmib='177
\newfam\cmmibfam
\textfont\cmmibfam=\tencmmib \scriptfont\cmmibfam=\sevencmmib
\scriptscriptfont\cmmibfam=\fivecmmib
\def\cmmib#1{{\fam\cmmibfam\relax#1}}
\numberwithin{equation}{section}
\def\a{{\eusm A}}
\def\b{{\eusm B}}
\def\neg{\negthinspace}
\def\d{\mathrm d}
\def\C{{\Bbb C}}
\def\HH{{\mathbb H}}
\def\P{{\mathcal P}}
\def\NS{{\sf{NS}}}
\def\Ra{{\sf{R}}}
\def\sV{{\sf V}}
\def\Z{{\Bbb Z}}
\def\A{{\eusm A}}
\def\B{{\eusm B}}
\def\S{{\mathcal S}}
\def\bar{\overline}
\def\sc{{\mathrm{sc}}}
\def\Max{{\mathrm{Max}}}
\def\CS{{\mathrm{CS}}}
\def\ga{\gamma}
\def\bg{\bar\ga}
\def\W{{\mathcal W}}
\def\M{{\mathcal M}}
\def\bM{{\overline \M}}
\def\L{{\mathcal L}}
\def\sM{{\sf M}}
\def\gst{\mathrm{g}_{\mathrm{st}}}
\def\gstt{\widetilde{\mathrm{g}}_{\mathrm{st}}}
\def\be{\begin{equation}}
\def\ee{\end{equation}}

\section{Introduction
\label{I}}
Genetic evolution \cite{Darwin,Mendel1,Mendel2,Mendel3} occurs through changes in the heritable characteristics of a population over successive generations. Individuals with traits that are advantageous for survival and reproduction are more likely to pass those traits to the next generation. Over time, these advantageous traits become more common in the population, leading to evolutionary changes. The other phenomena that could lead to evolutionary change is the mutaion which only changes a genetic code. It happens when a creature undergoes unexpected natural stresses and then it naturally selects the more stable state. It strongly correlates with external environmental conditions such as air pressure, temperature, food resources, cosmic rays, etc. It also depends on the occurrence of natural phenomena like earthquakes, asteroids, and hurricanes. 

The basic question one can ask is on what scale does evolution happen? We want to discuss the subject on a special scale where the quantum dynamic is dominant. If the evolution happens quantum mechanically, then the phase space describing the dynamics is the {\it{space of possible states}} or {\it{Hilbert space}} which contains finite possible states to be observed by the observer. The observed states are our observables in daily life but this does not mean that other states do not exist. This is due to the fundamental nature of {\it{uncertainty principle}}. This scenario points out that a cell can choose between an infinite number of possible states with random distribution each with a specific probability weight. We consider these states to be chosen from a hermitian random matrix ensemble by the use of Random Matrix Theory (RMT). This theory is a strong and useful tool to classify and describe the energy spectrum of many-body chaotic quantum systems with strong coupling between its partitions which are usually thermal systems. It has been intended to study statistical behaviors in quantum regime. Its applications are widely used in various fields, such as string theory, two-dimensional gravity \cite{Ginsparg-Moore,Francesco}, RNA folding \cite{RNA-folding}, etc. For useful reviews on this context see \cite{RMT1,RMT2,RMT3,RMT4}.

Evolution can be observed as changes in the ensemble of genes. The state of a living cell before the decomposition can be considered as a variable taken from a random matrix ensemble with Gaussian distribution. When the decomposition starts, it will pick other random variables from $n$ possible set of ensemble samples by a phase transition. This is what we call in this paper mutation and it does not always lead to an evolutionary change. If the order of matrices is $N\times N$, where $N\rightarrow \infty$, and if one consider the DNA as a two-dimensional lattice evaluates such that it goes from a weaker state to a stronger one, then this phase transition is very similar to the one proposed by \cite{Gross-Witten,Wadia} for QCD. By weak and strong we mean the stability of the cell against external effects. Just if the environmental conditions change intensely the decoupling may get a genetic disorder after the decomposition as one expects from a strongly coupled chaotic system. We will explain how this could be possible through the Gross-Witten-Wadia (GWW) phase transition by defining an effective coupling constant $g$ for DNA. The phase transition occurs at double-scaled limit $g\rightarrow g_c$ and $N\rightarrow \infty$ leaving 'tHooft parameter $\tau=g^2 N$ fixed where $g_c$ is a critical coupling, depending on the conditions of the system.

We will discuss cellular fission and compare it with a Gaussian distribution. The outcome state from the decomposition is picked randomly between many possible states on a random ensemble with a given probability. The main goal of the paper is the possible phase transition in RMT and its relation to the mutation, for a general set of random matrix distributions. Depending on the symmetries of the system, these ensembles could be Gaussian Unitary Ensemble (GUE), Gaussian Orthogonal Ensemble (GOE), Gaussian Symplectic Ensemble (GSE), or other class of distributions \cite{RMT3}. We will discuss the case of GUE.


\section{RMT, Phase Transition and Evolution
\label{III}}

\begin{figure}[!htbp] 
\begin{center}
\includegraphics[width=6.5cm,clip]{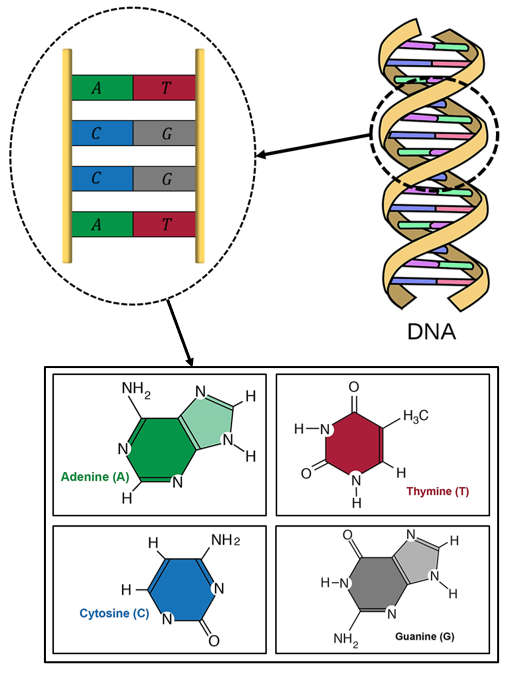}
\caption{The structure of DNA (up, right), the nucleotide base pairs (up, left), and the chemical structure of the fundamental nucleotides (down).}
\label{DNA}
\end{center}
\end{figure}

A single DNA is constructed of two {\it{sugar phosphate backbones}}, four {\it{nucleotides}} 
(see fig. \eqref{DNA}) and two AT and GC {\it{base pairs}}. GC content of a DNA segment, which is the percentage of nitrogenous bases in a DNA, can affect the stability of the DNA molecule. GC base pairs are held together by three hydrogen bonds, compared to two AT base pairs, making GC-rich DNA more thermally stable. This stability plays a role in the structure and function of genomic DNA, influencing aspects like gene expression and chromatin organization.

\vspace{0.2cm}

\begin{figure}[!htbp] 
\begin{center}
\includegraphics[width=7cm,clip]{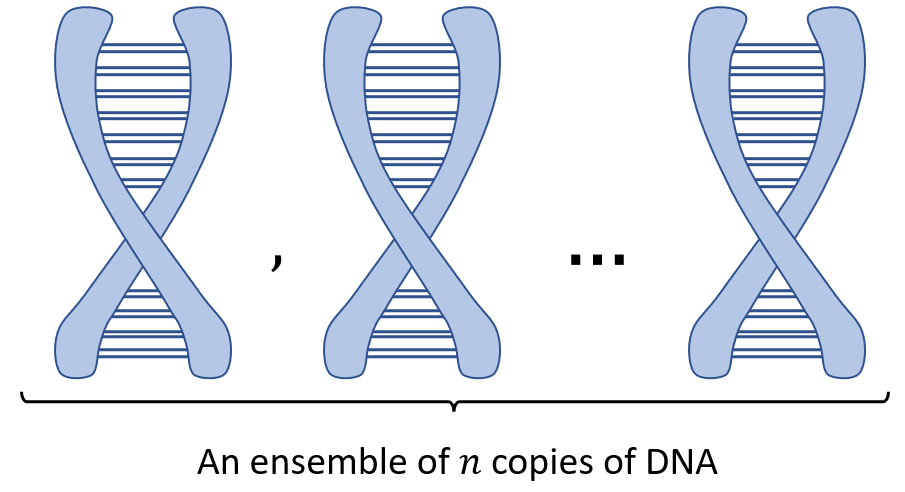}
\caption{Here we consider an $n$ copy of DNA sequences each with $N$ pairs, as samples of a random matrix ensemble. Each sample represents AT and GC base pairs with random energy eigenvalues taken from an $N\times N$ matrix \eqref{diagH} with positive and random entries.}
\label{Ncopies}
\end{center}
\end{figure}

We consider that each AT or GC base pair in the DNA represents a force on a specific distribution with zero mean and determined variance $\mathcal{V}$. These forces are related to the potential energy $F=-\nabla U$, therefore each covalent bond carries an amount of potential energy. In DNA, the energy can differ from one bond to another. For example, an AT bond on one part of the DNA can have slightly different energy than the AT bond on the other part. That is possible since this energy depends on the location of the nucleotides and many other side effects of the potentials of other objects. By seeing this bond energy as a random variable, get picked by the time from a Gaussian disterbution, the DNA can have statistical distribution in terms of energy eigenvalues.
\vspace{-0.4cm}\\ One can define a theory with Hamiltonian $H$ as a unitary matrix with random entries of the size of the number of DNA bonds $N$, in which the energy eigenvalues show the amount of the potential energy of each pair in each generation step in the whole ensemble. One can find the energy eigenvalues of each cell's DNA by diagonalization of H:
\begin{align}
diag(H)=\begin{pmatrix}
\lambda_{\text{AT}}^{(1)}&0&0&\dots~\\[0.2cm]
0&\lambda_{\text{GC}}^{(2)}&0&\dots~\\[0.2cm]
0&0&\lambda_{\text{GC}}^{(3)}&\dots~\\[0.2cm]
0&0&0&\dots~\\
\end{pmatrix}_{N\times N}
\label{diagH}
\end{align}
Where $\lambda_{\text{AT}}^{(i)}$ and $\lambda_{\text{GC}}^{(i)}$ is the energy of AT and GC for the $i$th pair respectively. Note that the structure of this theory is unknown and we are only extrapolate the general results by considering a system with such Hamiltonian. The ensemble is made of a large number of such matrices known as {\it{samples}}. Another picture would be an $n$ copy of DNA sequences (see fig.\eqref{Ncopies}). The values of entries in each sample could follow Gaussian or other random distributions. The randomness and the statistical behavior would be manifest at the limit $N\rightarrow \infty$. We discuss a hermitian system in this limit with a unitary matrix of Gaussian distribution.
\vspace{-0.4cm}\\
For such a system with finite and constant coupling $g$, there is Columb-like force that causes the eigenvalues to repel at short energy spacings and a {\it{long-range spectral rigidity}} at large energy spacings \cite{RMT2}. In genetics, the bonds also have some similar correlations among each other and this shows that both systems may appear to have the same universal features. For example, In the context of bacterial genomes, a study has shown that there is often a negative correlation between AT and GC pairs, meaning that as the proportion of AT base pairs increases, the proportion of GC base pairs tends to decrease, and vice versa. This correlation can vary among different organisms and can be influenced by factors such as the replication mechanisms of DNA \cite{corr-eig}.

\begin{figure}[!htbp] 
\begin{center}
\includegraphics[width=5cm,clip]{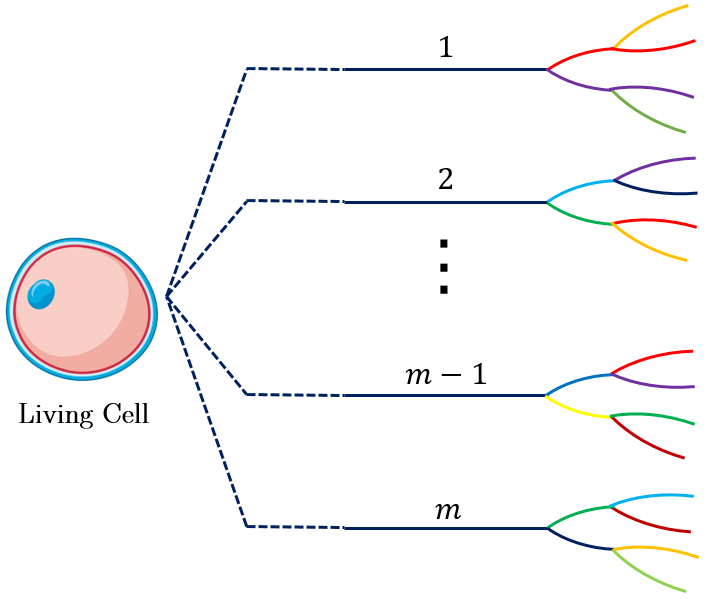}
\caption{If the mechanism of decomposition of the living cell is assumed to be quantum mechanical then there are $m$ possible states for the cell to choose. Each has its binary growth and probability weight. The states $1, 2, ..., m$ can be represented by an RMT ensemble with a Gaussian distribution. All states can be thought of as an ensemble of $n\gg1$ copies of DNA each with $N$ base pairs.}
\label{natural-selection}
\end{center}
\end{figure}

In cellular fission, the DNA will separate into two branches to construct two other cells from the same gene. In the whole process, the matrix elements will be generated randomly, each bond in a specific variance, and that will lead to completely different values for base pairs, for both produced cells. The cell will randomly choose one of the $m$ possible states (see fig. \eqref{natural-selection}) to start the replication process. Each state has its probability weight and the one which has greater probability is more likely to be {\it{naturaly selected}}.
\newpage
\subsection{GWW phase transition}
The main question of this paper is when genetic evolution occurs in the decomposition process. Before answering this question, let us see when evolution happens from a biological point of view. As a naive expression, When the cell starts to decouple into two individual cells, most of the time it will preserve the original gene. The story becomes interesting when one nucleotide (like A) exchanges with another nucleotide (like C) and causes an error in the replication which in turn changes the features of the living cell. Now the change $A\rightarrow C$ or $T\rightarrow G$ can change the energy levels of DNA spectrum (a gap on the Gaussian distribution) to put the cell in a more stable state (since GC pairs are stronger). This means one eigenvalue has changed the statistics of RMT which can slightly deviate the shape of the density of eigenvalues $\rho(\lambda)$ of that disterbution.

This evolution can be related to the {\it{phase transition}} in RMT since it describes a variation between two phases, weak (more AT) and strong (more GC). This is also possible through studying the distribution of the maximum eigenvalue or {\it{Tracy-Widom}} distribution \cite{TW1,TW2,TW3}. This has been widely studied in \cite{Majumdar2} and we are not going to go through the complete details in this paper. To support this idea, It has been argued in \cite{R-May} that any random complex interacting system will remain stable\footnote{The stability here means the perseverance of the species against external conditions. The system is unstable when the species goes toward extinction or the system is not prepared to preserve its heredity by evolution. Natural selection is an action against this process.} {\it{if}} the largest eigenvalue $\lambda_{\text{max}}$ of the random distribution is less than an inverse of the coupling constant of the system $g$:
\begin{align}
\lambda_{\text{max}}<(g)^{-1}
\end{align}
there is a critical value for $g$ in which at that point the system would experience a phase transition. A similar transition occurs in two-dimensional large-$N$ lattice gauge theories\footnote{For a useful review of large-N gauge theories see \cite{Marino}.} in critical coupling constant which is known as Gross-Witten-Wadia phase transition \cite{Gross-Witten,Wadia}. Using the {\it{steepest-descent}} method \cite{steepest-descent}, \cite{Gross-Witten} solved the large $N$ limit of the two-dimensional lattice gauge theory and observed a possible 3rd order phase transition. They showed for a general unitary RMT, the behavior of the density of eigenvalues for weak and strong coupling regimes is described by different analytical functions \cite{Gross-Witten}:
\begin{equation}
\rho(\lambda)=
\begin{cases}
\frac{2}{\pi \tau}cos\frac{\lambda}{2}\left(\frac{\tau}{2}-sin^2(\frac{\lambda}{2})\right)^{1/2}, & \tau\leq 2\\
\frac{1}{2\pi}\left(1+cos \lambda\right), & \tau=2\\
\frac{1}{2\pi}\left(1+\frac{2}{\lambda}cos \lambda\right), & \tau\geq 2
\label{disterbutions}
\end{cases}
\end{equation}
a phase transition is observed (for very large N) at $\tau=2$, marking a boundary between weak and strong coupling regimes. For $|\lambda|\leq \sqrt{2\tau}$ and $\tau \rightarrow 0$, the distribution is given by Wigner's semi-circle law:
\begin{align}
\rho(\lambda)\simeq\frac{1}{\pi}\sqrt{1-\lambda^2}
\label{wigner-semi-circle}
\end{align}
The phase transition takes place exactly when the eigenvalues completely occupy the entire circle. When the value of the coupling is significantly large, the theory would induce a repulsion among the eigenvalues, leading to a uniform distribution. For very small coupling, the distribution will be as random as possible. A string model \cite{t-Hooft-large-N} describing the large $N$ perturbative expansion of QCD could also arise within the lattice framework of this theory. Adjusting the lattice spacing $a$ and the coupling constant $g$ in such a way that the string tension remains constant would help to define an {\it{effective coupling}}. Using this coupling, \cite{Gross-Witten} showed that the $\beta$-function of the theory does not vanish at $\lambda=2$, meaning that this phase transition is of a higher order than two, otherwise the string tension must vanish. It is of third order since the third derivative of the free energy experiences a discontinuity at $\lambda=2$.

\begin{figure}[!htbp] 
\begin{center}
\includegraphics[width=9cm,clip]{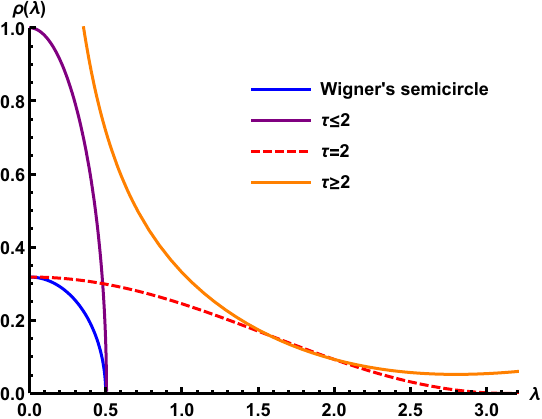}
\caption{The density of eigenvalues for \eqref{disterbutions} and \eqref{wigner-semi-circle} is plotted. Semi-circle law and small coupling distributions are drawn for $\tau=0.125$. For very small couplings, the semi-circle and $\tau\leq 2$ distributions almost behave the same close to $\lambda_{\text{max}}$ which is a sharp edge with many random values. As the coupling gets bigger, the distribution becomes uniform until it reaches $\tau=2$ where the phase transition occurs.}
\label{plot}
\end{center}
\end{figure}
Similarly, for a DNA system with random entries for bindings chosen from a unitary random matrix ensemble of $H$. We consider each DNA as a 2d lattice that its interactions are mainly electron-electron random columb forces, to be analogous to a $U(N)$ lattice gauge theory. In analogy to the string model, we also consider the bindings as strings with finite tension $\kappa$ with an average spacing $\bar{a}$. One can adjust the average spacing $\bar{a}$ and the coupling $g$ in such a way that the tension of each base pair remains fixed. Therefore we define an effective coupling $g(\bar{a})$ for DNA bindings such that $\kappa[ \bar{a}, g(\bar{a}) ]=\kappa$.

When the cell decouples in several steps, the structure of DNA may encounter some disorders from external effects that put the life of the cell at risk of distruction or cause an error in the replication. Therefore, the nucleotides will encounter instabilities and choose a stronger structure for the next steps of decoupling. Hence one of the nucleotides, for instance, A, exchanges with other ones that provide more powerful binding, and the DNA will turn into a more stable phase. The coupling constant does not always change in that system just if the system undergoes certain circumstances and tries to preserve its tension. Since there is an obvious interaction between nucleotides, this change in A, transforms the coupling constant to a new value which just at the transition phase, reaches its critical value $g_c$.
At the moment of decomposition, one specific binding should not necessarily have the exact value of previous energy and it gets picked randomly in later steps \footnote{To see this, one can also write the density of energy eigenvalues $\rho(\lambda)$ and $\rho(\tilde{\lambda})$ in a Gaussian ensemble and calculate the pair correlation function $\rho_{c}(\lambda,\tilde{\lambda})=\langle\rho(\lambda)\rho(\tilde{\lambda})\rangle$ and take the limit $\lambda-\tilde{\lambda}\rightarrow0$.}. We assume that the amount of energy of each pair $\lambda_{\text{AT}}$ and $\lambda_{\text{GC}}$ can not exceed a minimum and a maximum value. If so, the binding would lead to the destruction of one part of the DNA and the tension vanishes which we are not interested in.
\\Statistically, if the number of AT and GC energy bindings is equal (with $\lambda_{\text{GC}}>\lambda_{\text{AT}}$), changing $A\rightarrow C$ will change the distribution from the equilibrium point. Therefore, the interaction between AT and GC will turn into a new regime with a less coupling constant between molecules since $\lambda\sim g^{-1}$. That means the larger the population of GC base pairs, the easier and more accessible it can interact with neighboring eigenvalues. In other words, we tune the coupling such that the structure remains stable. 
This is similar to the GWW phase transition from weak to strong coupling phase which can turn the DNA into a more stable regime. Since this phase transition occurs randomly in the fission process, it is somehow equivalent to mutation.

\section{Conclusion}
It has been explained that the cell can select a state in which the base pair bindings take random and independent energy values in each time step of cellular fission. This is just one state between many other possible states in the ensemble. Each cell has many states to choose and evaluate, so there are an infinite number of states, each with its probability weight. Environmental parameters such as temperature can play a role in the probability value. We can conclude that the behavior of the cell is a random chaotic phenomenon that can be described in quantum mechanical viewpoints by random matrix theory in which the mutation is equaivalent to GWW phase transition. Through the idea that has been presented here, one can probe the existence of such transition by analyzing the long time statistics of a large number of samples for a specific DNA. This is directed for a future research project. The change in DNA bindings can directly change the overall coupling constant which leads to a GWW phase transition from unstable to stable regime and vice versa.
\\It is important to calculate the free energy of DNA in large $N$ so that it can be compared with the main results obtained for the string model \cite{t-Hooft-large-N}. 
\\Connecting the effects and mechanisms inside the cell to the mathematical developments of the random matrix theory, can take a great step in tracking the exact location and the time of mutations and drawing a historical map of different evolutions. By analyzing the spectrum and calculating the energy change in the GWW phase transition one expects to see in what exact statistical circumstances the cells have more chance to survive. 
\vspace{1cm}\\
{\bf Acknowledgments}: The author would like to thank M. Siahvoshan for her valuable comments.


\bibliographystyle{unsrt}

\end{document}